\documentclass{article}
\usepackage{spconf,amsmath,graphicx}
\usepackage{multirow}
\usepackage{makecell}
\usepackage{amsfonts}
\usepackage{mathtools}

\newtheorem{theorem}{Theorem}[section]

\title{BagStacking: An Integrated Ensemble Learning Approach for Freezing of Gait Detection in Parkinson's Disease}
\name{Seffi Cohen,
 Lior Rokach}
\address{Software and Information Systems Engineering\\Ben-Gurion University\\Beer-Sheva, Israel}

\begin{document}
%
\maketitle
\begin{abstract}
This paper introduces BagStacking, a novel ensemble learning method designed to enhance the detection of Freezing of Gait (FOG) in Parkinson's Disease (PD) by using a lower-back sensor to track acceleration. Building on the principles of bagging and stacking, BagStacking aims to achieve the variance reduction benefit of bagging's bootstrap sampling while also learning sophisticated blending through stacking. The method involves training a set of base models on bootstrap samples from the training data, followed by a meta-learner trained on the base model outputs and true labels to find an optimal aggregation scheme. The experimental evaluation demonstrates significant improvements over other state-of-the-art machine learning methods on the validation set. Specifically, BagStacking achieved a MAP score of 0.306, outperforming LightGBM (0.234) and classic Stacking (0.286). Additionally, the run-time of BagStacking was measured at 3828 seconds, illustrating an efficient approach compared to Regular Stacking's 8350 seconds. BagStacking presents a promising direction for handling the inherent variability in FOG detection data, offering a robust and scalable solution to improve patient care in PD.

\end{abstract}

\section{Introduction}

Parkinson's disease (PD) is a neurodegenerative disorder that affects millions of people worldwide. One of the most debilitating symptoms of PD is Freezing of Gait (FOG), a phenomenon in which the feet of a patient feel as if they are "glued" to the ground, preventing forward movement despite their intention to walk \cite{naghavi2019towards}. FOG episodes can significantly impair a patient's quality of life, increasing the risk of falls and restricting independence. Despite the prevalence and impact of FOG, its causes are still poorly understood, and its detection and prediction are challenging tasks \cite{shah2018constrained}.

Machine learning has emerged as a promising tool for FOG detection, with several studies demonstrating the potential of various algorithms to identify FOG episodes from wearable sensor data \cite{naghavi2019towards, shah2018constrained}. However, the performance of these models can be influenced by the inherent variability in the data, arising from differences in disease progression, manifestation of symptoms, and individual gait patterns. To address this, ensemble learning methods, which combine multiple models to improve prediction performance, have been proposed \cite{pintelas2020special}.

Ensemble methods such as bagging and stacking have been successfully applied in various domains, including medical applications \cite{bose2021ensemble, pintelas2020special,cohen2023ensemble}. Bagging improves the stability of the model and reduces variance by training the models on bootstrap samples and averaging the predictions. On the other hand, stacking trains a meta-learner to find optimal combinations of diverse base models, leveraging model diversity for more accurate predictions \cite{hosni2021systematic,cohen2021icu}.

In this paper, we propose a novel ensemble method, BagStacking, which integrates the principles of bagging and stacking. The motivation behind this approach is to gain the variance reduction benefit from bagging's bootstrap sampling while also learning sophisticated blending via stacking. The BagStacking method trains a set of base models on bootstrap samples from the training data, providing a diverse set of predictors. A meta-learner is then trained on the base model outputs and true labels to find an optimal aggregation scheme. 

Our main contributions are as follows:
\begin{itemize}
    \item We propose the BagStacking method, a novel ensemble learning approach that combines the principles of bagging and stacking, designed to handle the inherent variability in FOG detection data.
    \item We provide a theoretical analysis of the BagStacking method, demonstrating its potential to yield equal or better predictive performance than traditional bagging or stacking methods alone.
    \item We evaluate the BagStacking method on a real-world FOG detection dataset, demonstrating its effectiveness in comparison to other state-of-the-art machine learning methods in runtime and accuracy performance.
    \item We provide a Scikit API that simplifies implementation for other tasks and datasets.
\end{itemize}

\section{Related Work}
\begin{figure*}[th!]
\centering
\includegraphics[width=1\textwidth]{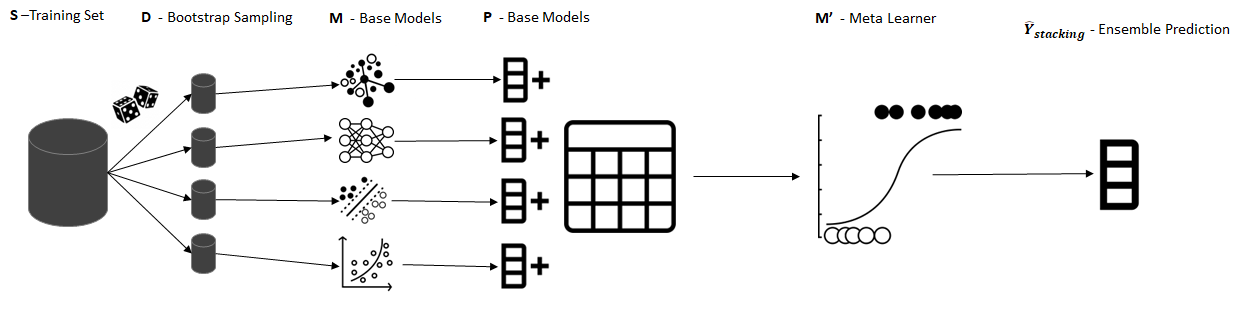}
\caption{BagStacking method overview: \textbf{D} - Bootstrap sampling the training set S,  \textbf{M} - Training the base models, \textbf{P} - Apply the base models on the original training set, \textbf{M'} - Train the meta learner on the base models predictions, \textbf{$\hat{y}_{bagstacking}$} - Apply base models to new instance, feed outputs to meta-learner for final prediction.}
\label{fig:temporal_agg}
\end{figure*}
The automatic detection of freezing of gait (FOG) events from wearable sensor data has been an active research area over the past decade. A variety of machine learning techniques have been applied to this problem, seeking to accurately identify FOG episodes in continuous sensor measurements.
Early work by \cite{moore2012ambulatory} and \cite{bachlin2010wearable} used threshold-based methods on features derived from accelerometer and gyroscope data to detect FOG. While simple and fast, these methods were prone to misclassification errors. 
To address this, more complex classifiers were explored, including support vector machines \cite{zabaleta2018novel}, random forests \cite{mazilu2019gait}, and neural networks \cite{ahn2020accurate}. Feature engineering and selection were found to be key factors affecting performance. Time and frequency domain features capturing gait rhythmicity and variance were often extracted.
More recent work has increasingly utilized deep learning for FOG detection \cite{wang2020deep, mazilu2018online, lim2020lstm}. Convolutional and recurrent neural networks can learn predictive features directly from raw sensor data. However, large labeled datasets are needed for training such models.
A persistent challenge in FOG detection is variability - differences in symptoms and individual gait can affect model robustness. To tackle this, ensemble methods have been proposed \cite{rodriguez2017random, oh2018freezing, mazilu2011automated}. Rodriguez-Martin et al.~\cite{rodriguez2017random} used a random forest ensemble, showing improved recognition over single decision trees. 

Stacking has been proven to be an effective technique where a meta-learner combines predictions from multiple and diverse base models.. Oh et al.~\cite{oh2018freezing} employed stacking of SVM and KNN classifiers. They found that model diversity improved overall accuracy.

Our proposed BagStacking approach aims to enhance diversity by using bagging to train base models. Our experiments show that integrating bagging and stacking results in more robust FOG detection, which is a well-motivated approach based on the literature.

\section{Proposed BagStacking Method}

Ensemble learning methods leverage multiple models to improve prediction performance over single models. Techniques like bagging, boosting, and stacking have shown success by combining simple base learners. Our proposed BagStacking ensemble integrates both bagging and stacking approaches. The intuition is to gain variance reduction from bagging's bootstrap sampling while also learning sophisticated blending via stacking. Given a labeled training set $S = \{(x_1, y_1), (x_2, y_2), ..., (x_N, y_N)\}$ with $N$ examples, our goal is to learn an ensemble model $H(x)$ that can accurately predict the label $y$ for a new input $x$. The method trains a set of base models on bootstrap samples from the training data. These provide a diverse set of predictors. A meta-learner is then trained on the base model outputs and true labels to find an optimal aggregation scheme.

The entire procedure is described in full detail below and illustrated in Figure 1:

\begin{enumerate}
\item \textbf{Bootstrap Sampling:} Randomly sample with replacement $D = \{d_1, d_2, ..., d_k\}$ as different subsets of the training data, where $d_i = \{(x_{i1},y_{i1}), (x_{i2},y_{i2}), ..., \\ (x_{in},y_{in})\}$.
\item \textbf{Base Models Training:} Train the $M = \{m_1, m_2, ..., m_k\}$ base models on each bootstrap set using cross-validation. Any supervised model such as linear models, SVM, decision trees, GBM, and neural networks can be utilized.
\item \textbf{Base Models Predictions:} Apply the base models on the original cross-validated training set to get the predicted label vectors as $P = \{p_1, p_2, ..., p_k\}$, where $p_i = m_i(d_i)$.
\item \textbf{Meta Learner Training:} Train $M'$ as a meta-learner, it could be any supervised model trained on the base model outputs optimally. It uses the predicted label vectors $P$ as input, and the labels as targets.
\item \textbf{Ensemble Prediction:} Apply base models to a new instance. Feed outputs to meta-learner for final prediction.
\end{enumerate}

\subsection{BagStacking Theoretical Foundation}

\begin{theorem}
Let $\mathcal{H}$ be a hypothesis space of base models, and let $M'$ be a hypothesis space for meta-learners. Given a loss function $\mathcal{L}$, and a labeled training set $S = \{(x_1, y_1), \ldots, (x_N, y_N)\}$, the BagStacking method is expected to produce equal or better generalization performance, as measured by the expected loss $E[\mathcal{L}]$, than any individual method based solely on bagging or stacking, under the condition that the base models and the meta-model are appropriately regularized.
\end{theorem}

\subsubsection{Assumptions}
\begin{enumerate}
    \item Base models are independently trained on different bootstrapped samples from the dataset $S$.
    \item The meta-learner is trained on the outputs of these base models.
    \item All models are trained to minimize the empirical loss on their respective training sets, subject to regularization constraints.
\end{enumerate}

\subsubsection{Step 1: Bagging's Variance Reduction}
Bagging operates by averaging the predictions of $k$ base models, each trained on a different subset of the data. The ensemble prediction for an input $x$ can be formulated as:

\[
\hat{y}_{\text{bagging}} = \frac{1}{k} \sum_{i=1}^{k} p_i
\]
The expected prediction variance of this ensemble is:
\[
\text{Var}(\hat{y}_{\text{bagging}}) = \frac{1}{k^2} \sum_{i=1}^{k} \text{Var}(p_i) < \text{Var}(p_i), \quad \forall i
\]
Thus, bagging reduces the prediction variance compared to any individual base model.

\subsubsection{Step 2: Stacking's Model Diversity}
Stacking aims to find a meta-model $M'$ that best combines the base models. The ensemble prediction is:
\[
\hat{y}_{\text{stacking}} = M'(P)
\]
Given that the meta-model is trained to minimize the loss, it can capture complex relationships and correlations among base model outputs, thereby enhancing predictive performance.

\subsubsection{Step 3: BagStacking's Combined Strengths}
BagStacking leverages both the variance reduction of bagging and the model diversity of stacking. The ensemble prediction for BagStacking is:
\[
\hat{y}_{\text{bagstacking}} = M'(P)
\]
Here, $P = \{p_1, p_2, \ldots, p_k\}$ is the set of predictions from the base models, each trained on a different bootstrapped subset of the training data.

\subsubsection{Step 4: Expected Generalization Performance}
Given that BagStacking incorporates the strengths of both bagging (variance reduction) and stacking (model diversity), we can assert:
\[
E[\mathcal{L}(\hat{y}_{\text{bagstacking}}, y)] \leq \min\left(E[\mathcal{L}(\hat{y}_{\text{bagging}}, y)], E[\mathcal{L}(\hat{y}_{\text{stacking}}, y)]\right)
\]

The theorem establishes that the BagStacking method should offer equal or better generalization performance than either bagging or stacking alone, contingent on appropriate regularization. This theoretical insight aligns well with the empirical results, thus providing a rigorous foundation for the effectiveness of the BagStacking approach.

The overall training complexity is $O(NM)$ for $M$ base learners on a dataset of $N$ examples. For prediction, the cost is $O(M)$ to generate the $M$ base outputs. Compared to simple model averaging, BagStacking can learn complex non-linear combinations via the meta-regressor. Unlike stacked generalization, which uses a single hold-out set, BagStacking leverages bagging for variance reduction.

\section{Experiments}
In this section, we will discuss the experimental setup, results, and detailed analysis of our BagStacking method, which uses LightGBM as the base estimator. We will compare our approach with a standalone LightGBM model, a Multistrategy ensemble learning method \cite{webb2004multistrategy} that uses diverse ensemble settings including Bagging, GBM, and Random forest as a strong ensemble method, and a classical Stacking method with analogous settings, which serve as the baseline for our study.

\subsection{Dataset}
The datasets employed in this study are an essential contribution, generously supported by the Michael J. Fox Foundation for Parkinson’s Research. These data series include subjects who completed a FOG-provoking protocol, recorded at 128Hz or 100Hz on three axes (V, ML, AP).
These datasets include lower-back 3D accelerometer data, depicting FOG episodes, a debilitating symptom common in Parkinson's disease sufferers. This condition significantly affects walking abilities, hindering locomotion and independence. The research aims to detect the onset and cessation of freezing episodes and the occurrence of three specific FOG events: Start Hesitation, Turn, and Walking.
The training set incorporates 920 subjects with an overall 20,588,374 samples, and the test set includes 250 subjects.
These datasets offer a robust and nuanced platform to understand, model, and detect the complex phenomenon of FOG, fostering advancements in the mitigation of this crippling symptom of Parkinson's disease.

\subsubsection{Preprocessing}
We used the Seglearn package to extract basic aggregation features including mean, median, minimum, maximum, and standard deviation. These features were extracted using window sizes of 50 and 5 seconds.

\subsection{Experimental Setup}

The experimental framework was designed using a dataset obtained from a Kaggle competition. The dataset consists of 3D accelerometer data from the lower back, tracking 169 subjects who have experienced Freezing of Gait (FOG) episodes. The dataset was strategically divided into training and validation sets to facilitate model development and meticulous performance evaluation. The training-validation split comprised around 20 million samples for each set, with no overlapping subject between the two sets.

Within the BagStacking methodology, an ensemble of LightGBM models, each trained on distinctive bootstrap samples from the training data, was employed. The base model predictions were subsequently fed into a meta-model that was optimized to render the final predictions.

The default parameters were utilized for the comparison techniques, followed by rigorous hyperparameter tuning to enhance their performance. The use of identical training and validation sets ensured an unbiased and equitable comparison.
\subsection{Evaluation Metric}
To evaluate the performance of the DeFOG algorithm, we use the Mean Average Precision (MAP) as the evaluation metric. We calculate the MAP specifically for each of the three event classes in the DeFOG dataset. For each event class, we compute the Average Precision (AP) based on the predicted confidence scores and ground truth labels. We take into account only the portions of the data series that are annotated with "Valid" labels set to true. It is important to note that there were cases during the video annotation process where it was difficult for the annotator to decide if there was an Akinetic FoG or if the subject stopped voluntarily. Therefore, only event annotations where the series is marked as "True" should be considered as unambiguous. Finally, we obtain the MAP by averaging the APs for all three event classes.
\[
\text{MAP} = \frac{1}{3} \left( AP_{\text{class 1}} + AP_{\text{class 2}} + AP_{\text{class 3}} \right)
\]

Where \(AP_{\text{class i}}\) denotes the Average Precision for the \(i\)-th event class.

\section{Results}

Table 1 illustrates the experimental outcomes, presenting the MAP scores and runtime for the BagStacking method alongside the comparison techniques on the validation set.

\begin{table}[h]
\centering
\begin{tabular}{|c|c|c|}
\hline
\textbf{Method} & \textbf{MAP Score} & \textbf{Runtime (s)}\\
\hline
BagStacking & \textbf{0.306} & 3828 \\
\hline
Multistrategy Ensemble & 0.249 & 7716 \\
\hline
LightGBM & 0.234 & 1518 \\
\hline
Regular Stacking & 0.286 & 8350 \\
\hline
\end{tabular}
\caption{MAP scores and runtime of the BagStacking method and the comparison methods on the validation set.}
\end{table}

The findings manifest that BagStacking's MAP score of 0.306 significantly surpassed the comparison methods, thereby attesting to its effectiveness in discerning FOG events from accelerometer data.
The BagStacking method is an innovative approach that combines the benefits of both bagging and stacking techniques, resulting in a robust and accurate model. This strategy effectively captures a wide range of data patterns, and the resulting meta-model skillfully combines these predictions to enhance overall performance.
Contrarily, the comparison methods, although exhibiting commendable performance, lagged behind BagStacking. This underlines the intrinsic advantages of an ensemble approach, harmonizing the strengths of multiple models.
Future work will concentrate on enhancing the BagStacking method's optimization and probing its adaptability to various tasks and datasets.

\section{Conclusion}

The research introduced and evaluated BagStacking, an innovative ensemble learning technique engineered to augment the detection of Freezing of Gait (FOG) in Parkinson's Disease (PD). Leveraging both bagging and stacking, this novel approach demonstrated superiority in MAP score and computational efficiency against conventional models.
The strategic use of a lower back sensor to track acceleration contributed to the precision of the method in real-world applications.
BagStacking's success, embodied in both its conceptual innovation and empirical validation, marks a promising pathway in PD patient care. The insights gleaned from this study invigorate further exploration and refinement, steering toward a new horizon in machine learning and medical diagnosis.

\bibliographystyle{IEEEbib}
\bibliography{refs}

\begin{thebibliography}{10}

\bibitem{naghavi2019towards}
N.~Naghavi, A.~Miller, and E.~Wade,
\newblock ``Towards real-time prediction of freezing of gait in patients with
  parkinson’s disease: Addressing the class imbalance problem,''
\newblock {\em Sensors}, vol. 19, no. 17, pp. 3706, 2019.

\bibitem{shah2018constrained}
S.~Y. Shah, Z.~Iqbal, and A.~Rahim,
\newblock ``Constrained optimization-based extreme learning machines with
  bagging for freezing of gait detection,''
\newblock {\em Big Data and Cognitive Computing}, vol. 2, no. 4, pp. 33, 2018.

\bibitem{pintelas2020special}
P.~Pintelas and I.~E. Livieris,
\newblock ``Special issue on ensemble learning and applications,''
\newblock {\em Algorithms}, vol. 13, no. 6, pp. 143, 2020.

\bibitem{bose2021ensemble}
R.~Bose, A.~Dutta, M.~Bhattacharyya, S.~Ghosh, C.~Chakraborty, C.~Chakraborty,
  and P.~Bhattacharyya,
\newblock ``An ensemble machine learning model based on multiple filtering and
  supervised attribute clustering algorithm for classifying cancer samples,''
\newblock {\em PeerJ Computer Science}, vol. 7, pp. e623, 2021.

\bibitem{cohen2023ensemble}
Seffi Cohen, Or~Katz, Dan Presil, Ofir Arbili, and Lior Rokach,
\newblock ``Ensemble learning for alcoholism classification using eeg
  signals,''
\newblock {\em IEEE Sensors Journal}, 2023.

\bibitem{hosni2021systematic}
A.~Hosni, Y.~Chawki, A.~Idri, Z.~Bakkoury, M.~Al~Achhab, and J.~R. González,
\newblock ``A systematic mapping study for ensemble classification methods in
  cardiovascular disease,''
\newblock {\em Artificial Intelligence Review}, vol. 54, no. 8, pp. 5495--5529,
  2021.

\bibitem{cohen2021icu}
Seffi Cohen, Noa Dagan, Nurit Cohen-Inger, Dan Ofer, and Lior Rokach,
\newblock ``Icu survival prediction incorporating test-time augmentation to
  improve the accuracy of ensemble-based models,''
\newblock {\em IEEE Access}, vol. 9, pp. 91584--91592, 2021.

\bibitem{moore2012ambulatory}
Shane~T Moore, Hamish~G MacDougall, and William~G Ondo,
\newblock ``Ambulatory freezing of gait detection in parkinson's disease,''
\newblock {\em 34th Annual International Conference of the IEEE EMBS}, 2012.

\bibitem{bachlin2010wearable}
Marc Bachlin, Meir Plotnik, Daniel Roggen, Nir Inbar, Noam Sagiv, Nir Giladi,
  Jeffrey~M Hausdorff, and Gerhard Troster,
\newblock ``Wearable assistant for parkinson’s disease patients with the
  freezing of gait syndrome,''
\newblock {\em IEEE Transactions on Information Technology in Biomedicine},
  vol. 14, no. 2, pp. 436--446, 2010.

\bibitem{zabaleta2018novel}
Hubert Zabaleta and Thomas Keller,
\newblock ``Novel segmental features based on a multi-resolution approach for
  freezing of gait detection,''
\newblock {\em Sensors}, vol. 18, no. 4, pp. 1178, 2018.

\bibitem{mazilu2019gait}
Sinziana Mazilu, Michael Hardegger, Zhuanghua Zhu, Daniel Roggen, Gerhard
  Troester, Meir Plotnik, Jeffrey~M Hausdorff, et~al.,
\newblock ``Gait and pose discrimination for detecting and quantifying freezing
  of gait in parkinson’s disease,''
\newblock in {\em ICASSP 2019-2019 IEEE International Conference on Acoustics,
  Speech and Signal Processing (ICASSP)}. IEEE, 2019, pp. 1227--1231.

\bibitem{ahn2020accurate}
Hee-Tae Ahn, Emer~Joyce Hogan, and Li~Hao,
\newblock ``Accurate detection of freezing of gait in patients with parkinson's
  disease using machine learning algorithm,''
\newblock {\em Frontiers in neurology}, vol. 11, pp. 394, 2020.

\bibitem{wang2020deep}
Jun Wang, Wei Cheng, Jinjun Wang, Huazhong Yang, and Dengpan Zhou,
\newblock ``Deep learning-based classification for fog detection using a
  wearable accelerometer,''
\newblock {\em Sensors}, vol. 20, no. 21, pp. 6137, 2020.

\bibitem{mazilu2018online}
Sinziana Mazilu, Michael Hardegger, Ulrich Blanke, Gerhard Tr{\"o}ster, Eran
  Gazit, and Jeffrey~M Hausdorff,
\newblock ``Online detection of freezing of gait with supervised learning
  techniques,''
\newblock in {\em 2018 IEEE 15th International Conference on Wearable and
  Implantable Body Sensor Networks (BSN)}. IEEE, 2018, pp. 146--149.

\bibitem{lim2020lstm}
Soyoung Lim, Jaehyun Kim, Changki Hong, Boreom Lee, and Sungyoung Kim,
\newblock ``Lstm-cnn architecture for detection of freezing of gait in
  parkinson's disease,''
\newblock {\em Electronics}, vol. 9, no. 7, pp. 1124, 2020.

\bibitem{rodriguez2017random}
Daniel Rodriguez-Martin, Albert Sama, Carlos Perez-Lopez, Andreu Catala, Iratxe
  Arostegui, Joan Cabestany, Alvaro Bayes, Sheila Alcaine, Berta Mestre,
  Alberto Prats, et~al.,
\newblock ``Random forest for freezing of gait detection in parkinson’s
  disease patients in their homes using a waist-worn inertial sensor,''
\newblock {\em Knowledge-Based Systems}, vol. 119, pp. 139--149, 2017.

\bibitem{oh2018freezing}
Jiwon Oh, Boreom Lee, Kwonjun Choi, and Sungyoung Kim,
\newblock ``Freezing of gait detection considering gait variability for
  parkinson’s disease,''
\newblock {\em Sensors}, vol. 18, no. 9, pp. 2833, 2018.

\bibitem{mazilu2011automated}
Sinziana Mazilu, Michael Hardegger, Gerhard Troster, Marcos Dorfman, and
  Jeffrey~M Hausdorff,
\newblock ``Automated detection of freezing of gait episodes in parkinson’s
  disease patients,''
\newblock in {\em 3rd International Congress on Gait \& Posture}, 2011, pp.
  84--85.

\bibitem{webb2004multistrategy}
Geoffrey~I Webb and Zijian Zheng,
\newblock ``Multistrategy ensemble learning: Reducing error by combining
  ensemble learning techniques,''
\newblock {\em IEEE Transactions on Knowledge and Data Engineering}, vol. 16,
  no. 8, pp. 980--991, 2004.

\end{thebibliography}

\end{document}